\documentclass[11pt,a4paper]{article} 
\usepackage{jcappub} 
\usepackage{subcaption}

\title{Black holes in a cubic Galileon universe}

\author[a]{E.~Babichev,} 

\author[a]{C.~Charmousis,}

\author[a]{A.~Leh\'ebel} 

\author[a,b]{and T.~Moskalets} 

\affiliation[a]{Laboratoire de Physique Th\'eorique, CNRS, Univ. Paris-Sud, \\ Universit\'e Paris-Saclay, 91405 Orsay, France}
\affiliation[b]{V. N. Karazin Kharkiv National University, Kharkiv, Ukraine}

\emailAdd{eugeny.babichev@th.u-psud.fr}
\emailAdd{christos.charmousis@th.u-psud.fr}
\emailAdd{antoine.lehebel@th.u-psud.fr}
\emailAdd{tetiana.moskalets@th.u-psud.fr}

\abstract{We find and study the properties of black hole solutions for a subclass of Horndeski theory including the cubic Galileon term. The theory under study has shift symmetry but not reflection symmetry for the scalar field. The Galileon is assumed to have linear time dependence characterized by a velocity parameter. We give analytic 3-dimensional solutions that are akin to the BTZ solutions but with a non-trivial scalar field that modifies the effective cosmological constant. We then study the 4-dimensional asymptotically flat  and de Sitter solutions. The latter present three different branches according to their effective cosmological constant. For two of these branches, we find families of black hole solutions, parametrized by the velocity of the scalar field. These spherically symmetric solutions, obtained numerically, are different from GR solutions close to the black hole event horizon, while they have the same de-Sitter asymptotic behavior. The velocity parameter represents black hole primary hair.}

\begin{document}

\maketitle


\section{Introduction}

In recent years, it has become clear that Horndeski theory~\cite{Horndeski74} allows for black hole solutions with non-trivial scalar field configurations, which in some cases may be referred to as primary or secondary ``hair". Analytic solutions were first obtained in \cite{blaise} via Kaluza Klein compactification of Lovelock black holes \cite{zegers} and for translation invariant Galileon theories in \cite{Rinaldi:2012vy}. Horndeski theory (dubbed Galileon in modern interpretation~\cite{Nicolis:2008in,Deffayet:2009wt,Deffayet:2011gz}) is the most general scalar-tensor theory with equations of motion that contain no more than two derivatives. 
The last requirement is sufficient to avoid the Ostrogradski instability~\cite{Ostrogradski},  associated with higher-order derivatives.
By construction, Horndeski theory contains kinetic scalar-tensor coupling, and therefore the theory has richer phenomenology than, say, General Relativity (GR) with a minimally coupled scalar field.

A no-hair theorem, nevertheless, has been established for the shift-symmetric Galileon model~\cite{Hui:2012qt}.
The proof is based on a concrete physical assumption: the norm of the Noether current, associated with the shift symmetry of the model, is finite on the black hole horizon.
Later, however, two important counter-arguments have been found; one involving the Gauss-Bonnet invariant, and the latter discarding the theorem altogether \cite{Babichev:2016rlq}.
In the former, the Gauss-Bonnet invariant sources the scalar field equation~\cite{Sotiriou:2013qea} yielding a non-trivial scalar configuration. In this case, the norm of the Noether current is divergent on the horizon, although it is unclear if this is a physical problem for the special case of the Gauss-Bonnet scalar.   
The latter argument is in fact suggested by the behavior of a shift-symmetric Galileon in a cosmological setup: 
generically, the time-derivative of the scalar field is constant (rather than trivial) for cosmological attractors, e.g. see a discussion in~\cite{Babichev:2011iz}. 
Therefore, the assumption that the Galileon field is static does not have physical ground anymore, in contrast to, for example, a canonical scalar sitting in a minimum of some potential. This is also true for self tuning cosmological backgrounds \cite{st} in Fab 4 theories \cite{fab4}.
One may thus naturally require that the Galileon scalar has time-dependent asymptotic behavior, 
since one of the main physical motivations to consider Horndeski theory is to explain Dark Energy.
Put precisely, a black hole solution should have the asymptotes corresponding to some cosmological solution of a particular Galileon model.

In~\cite{Babichev:2013cya}, such an idea has been realized, allowing the scalar field to depend on time, while keeping a static metric. The theorem is then redundant since the field equations themselves dictate regularity of the Noether current and a non trivial scalar field \cite{Babichev:2016rlq}.  The full class of solutions has been found for a particular Galileon model, containing the ``John'' term, by classification of~\cite{fab4}.
Such a construction is not reserved to the presence of the ``John'' term. As it has been shown later~\cite{EBCCMH15}, the ansatz used in~\cite{Babichev:2013cya} leads to a consistent system of ODEs, i.e. the number of independent variables is equal to the number of equations. {\it Per se} this does not guarantee the existence of a solution, but shows self-consistency of the method. 
Indeed, in a number of works~\cite{Bravo-Gaete:2013dca,Kobayashi:2014eva,Charmousis:2014zaa}, other black hole solutions with a time-dependent Galileon have been found.

In this paper, we follow the method suggested in~\cite{Babichev:2013cya,EBCCMH15} to study black hole solutions in the shift-symmetric theory entailing the cubic Galileon term. Our motivation is three-fold. First, the technique for constructing black hole solutions of Refs.~\cite{Babichev:2013cya,EBCCMH15} has been applied to a particular Lagrangian, whose higher-order derivative part is of the ``John'' type.
Although later it was generalized to black hole solutions for a larger group of Lagrangians which have reflection symmetry~\cite{Kobayashi:2014eva}, the question is still unsettled for theories without this symmetry, see e.g. a comment on this point in~\cite{Maselli:2015yva}.
Secondly, the cubic Galileon can be viewed as the simplest Galileon with higher-order derivatives. It arises in various contexts, e.g. in the well-known Dvali-Gabadadze-Porrati (DGP) brane model~\cite{Dvali:2000hr}, as a particular limit~\cite{NR04}.
The third reason is that the cubic Galileon has been extensively studied in the cosmological context~\cite{Deffayet:2010qz} as dark energy with well behaved perturbations, and -- for the same model -- in the context of local Solar system physics~\cite{EBGEF13}.

The paper is organized as follows. In Sec.~\ref{Sec:setup} we start with the Lagrangian, equations of motion and the ansatz. 
Then, in Sec.~\ref{Sec:3D}, as a warm up, we study solutions for black holes in the cubic Galileon model in 3D. 
The equations of motion in 3D allow for analytic black hole solutions with Ba$\tilde{\text{n}}$ados-Teitelboim-Zanelli (BTZ) metric and nontrivial scalar configuration, 
which can be interpreted as secondary hair. 
Section~\ref{Sec:4D} is devoted to analytic properties of black hole solutions in 4D and Sec.~\ref{sec:Numerics} to the numerical integration of the field equations and subsequent analysis of the solutions. We conclude in Sec.~\ref{Sec:con}.

\section{Setup: action, equations of motion and ansatz}
\label{Sec:setup}

Throughout the paper we consider the following action:
\begin{equation}
S = \displaystyle\int{\mathrm{d}^Dx \sqrt{-g} \left[\zeta\: (R -2\: \Lambda)- \eta\: (\partial \phi)^2 + \gamma\: \Box \phi\: (\partial \phi)^2  \right]},
\label{eq:action}
\end{equation}
where $D$ is the number of dimensions (we will consider 3- and 4-dimensional cases), 
$\zeta$, $\eta$, $\gamma$ and $\Lambda$ are constant parameters of the Lagrangian. 
The third term in (\ref{eq:action}) is the DGP-like non-canonical Galileon term~\cite{NR04} 
and $\Lambda$ is the bare cosmological constant.

The variation of (\ref{eq:action}) with respect to the metric gives
\begin{equation}
\begin{split}
-\zeta (G_{\mu \nu} +\Lambda \: g_{\mu \nu}) - \eta \left[\dfrac{1}{2} g_{\mu \nu} (\partial \phi)^2 - \partial_\mu \phi \: \partial_\nu \phi \right] +\\
 +\gamma \left[- \Box \phi \: \partial_\mu \phi \: \partial_\nu \phi + \partial_{(\mu} \phi \: \partial_{\nu)} (\partial \phi)^2 \vphantom{\left[(\partial \phi)^2\right]} 
-\dfrac{1}{2} g_{\mu \nu} \partial^\rho \phi \: \partial_\rho \left[(\partial \phi)^2\right] \right] = 0.
\label{eq:EOM1}
\end{split}
\end{equation}
Since the action (\ref{eq:action}) is shift-symmetric, i.e. it is invariant under the transformation $\phi\to\phi +constant$, 
the scalar equation of motion can be written in terms of a conserved current $J^\mu$:
\begin{equation}\label{Jcons}
\nabla_\mu J^\mu =0,
\end{equation}
where the current associated with the scalar shift-symmetry is obtained by variation of (\ref{eq:action}) with respect to $\partial_\mu\phi$:
\begin{equation}
J^\mu =\partial_\nu \phi \left[g^{\mu\nu}(\gamma \Box \phi-\eta) -\gamma \nabla^\mu \nabla^\nu \phi \right]
\end{equation}

We assume a spherically symmetric ansatz for the metric and time-dependent for the scalar field, see~\cite{Babichev:2013cya}:
\begin{equation}
\begin{split}
\phi(t,r) &= q\, t + \displaystyle\int{\mathrm{d}r \: \dfrac{\chi(r)}{h(r)}},\\
\mathrm{d}s^2 &= - h(r) \mathrm{d}t^2 + \frac{\mathrm{d}r^2}{f(r)} + r^2 \mathrm{d}\Omega_{D-2}^2,
\end{split}
\label{eq:ansatz}
\end{equation}
where $q$ is a constant parameter that we call velocity, and $h(r)$ inside the integral is introduced for convenience. 

With the ansatz~(\ref{eq:ansatz}), for shift-symmetric Lagrangians, the equation $J^r = 0 $ is equivalent to the $(t r)$ component of the 
metric equations~\cite{EBCCMH15} for $q\neq 0$. We note that, apart from the solution $\phi'=0$, we can also have a non-trivial configuration for the scalar field $\phi$. Let us see explicitly how this comes about.

\section{Black holes in 3D}
\label{Sec:3D}

In three dimensions, subsituting our ansatz~(\ref{eq:ansatz}) yields the following equations of motion:
\begin{eqnarray}
\gamma \: q \: (r^2 h)'  \left(\dfrac{f}{h} \chi^2-q^2 \right) +2 \: \gamma \: q^3 \: r h - 2\: \eta \: q \: r^2 h \chi &=& 0, \label{eq:tr3d}\\
\eta \: r \left(\dfrac{f}{h} \chi^2 - q^2 \right) + \zeta \: f h' + 2\: \zeta\Lambda \: h r  &=& 0, \label{eq:mix13d}\\
\left(\dfrac{f}{h} \chi^2 - q^2 \right) \left[ \eta \: r \sqrt{\dfrac{h}{f}} - \gamma \: \left(r \sqrt{\dfrac{f}{h}} \chi \right)' \right] &=& \zeta \: h^2 \left(\sqrt{\dfrac{f}{h}} \right)',
\label{eq:mix23d}
\end{eqnarray}
%
%
where a prime denotes a derivative with respect to $r$. 
Equation (\ref{eq:tr3d}) is the $(t r)$ metric equation (or, equivalently, $J^r=0$). 
Equations~(\ref{eq:mix13d}) and~(\ref{eq:mix23d}) are a combination of the $(t r)$, $(t t)$ and $(r r)$ metric equations. 
The $(\theta \theta)$ and $(\varphi \varphi)$ metric equations are redundant, due to the Bianchi identities. 
The scalar field equation $\nabla_\mu J^\mu =0$ is also redundant, since $J^r=0$, \cite{EBCCMH15}. 
This ensures that the system of equations is not over constrained: we have three independent equations for three unknown functions. 
Let us note the structure of these equations: Eq.~(\ref{eq:tr3d}) is a second order polynomial equation in terms of $\chi$. 
Equation~(\ref{eq:mix13d}) is an algebraic equation in terms of $f$. 
If $\chi$ is substituted from the $(t r)$ equation, it is of third order, though it is not evident in this form.


We now assume additionally that $f = h$. 
The equations of motion are then easy to integrate, and we find the following solution:
\begin{equation}
\begin{split}
h(r) &= f(r) = - M  + \dfrac{\eta^2}{4 \: \lambda_\pm \: \gamma^2} r^2 ,\\
\chi(r) &= \dfrac{\eta r}{2 \gamma},
\end{split}
\label{sol3D}
\end{equation}
where 
$$
\lambda_\pm = \frac{\zeta \eta^2}{-2 \: \zeta \Lambda \: \gamma^2 \pm \displaystyle\sqrt{-2 \: \gamma^2 \: \zeta \: \eta^3+4 \:\gamma^4 \: \zeta^2 \Lambda^2}},
$$ 
and $M$ is an integration constant, corresponding to the mass of the black hole for $M > 0$. In the above solution, the velocity $q$ is a function of the Lagrangian parameters and the mass $M$. An event horizon exists as long as $\lambda_\pm>0$, and we are locally in anti de Sitter space. This family of solutions therefore exhibits the same behavior as the BTZ solution \cite{BTZ92}, with an effective cosmological constant $\Lambda_\mathrm{eff} = - \eta^2/(4 \lambda_\pm \gamma^2)$.

However, unlike the BTZ solution, (\ref{sol3D}) contains a nontrivial scalar field (see also \cite{stealth}) which effectively modifies the cosmological constant. 
The solution admits secondary hair, as it depends only on the mass of the black hole. 
The different possibilities for the effective cosmological constant can be parametrized in the following way, 
which will also prove useful in four dimensions: 
\begin{equation}
\Lambda_\mathrm{eff}=
\left\{
\begin{array}{r c l}
\Lambda_< &=& \dfrac{1}{2} \left(\Lambda- \sqrt{\Lambda^2+3\: \Lambda_\mathrm{KGB}^2}\right) \ \text{if} \ \eta <0\\
\\
\Lambda_>^+ &=& \dfrac{1}{2} \left(\Lambda + \sqrt{\Lambda^2-3\: \Lambda_\mathrm{KGB}^2}\right) \ \text{if} \ \eta >0 \ \text{and} \ |\Lambda | > \sqrt{3} \: \Lambda_\mathrm{KGB}\\
\\
\Lambda_>^- &=& \dfrac{1}{2} \left(\Lambda- \sqrt{\Lambda^2-3\: \Lambda_\mathrm{KGB}^2}\right) \ \text{if} \ \eta >0 \ \text{and} \ |\Lambda| > \sqrt{3} \: \Lambda_\mathrm{KGB}
\end{array}
\right.
\label{eq:Lambdaeff3D}
,\end{equation}
where we assume $\zeta>0$ and we introduce the parameter
\begin{equation}\label{LambdaKGB}
\Lambda_\mathrm{KGB}  \equiv \left(\dfrac{|\eta|^3}{6 \zeta \gamma^2}\right)^{1/2},
\end{equation}
which, in 4D (see below), is connected to the effective cosmological constant in the case of ``Kinetically Braiding Gravity''~\cite{Deffayet:2010qz}, hence the subscript KGB. 

\begin{figure}[t]
\begin{center}
\includegraphics[width=10cm]{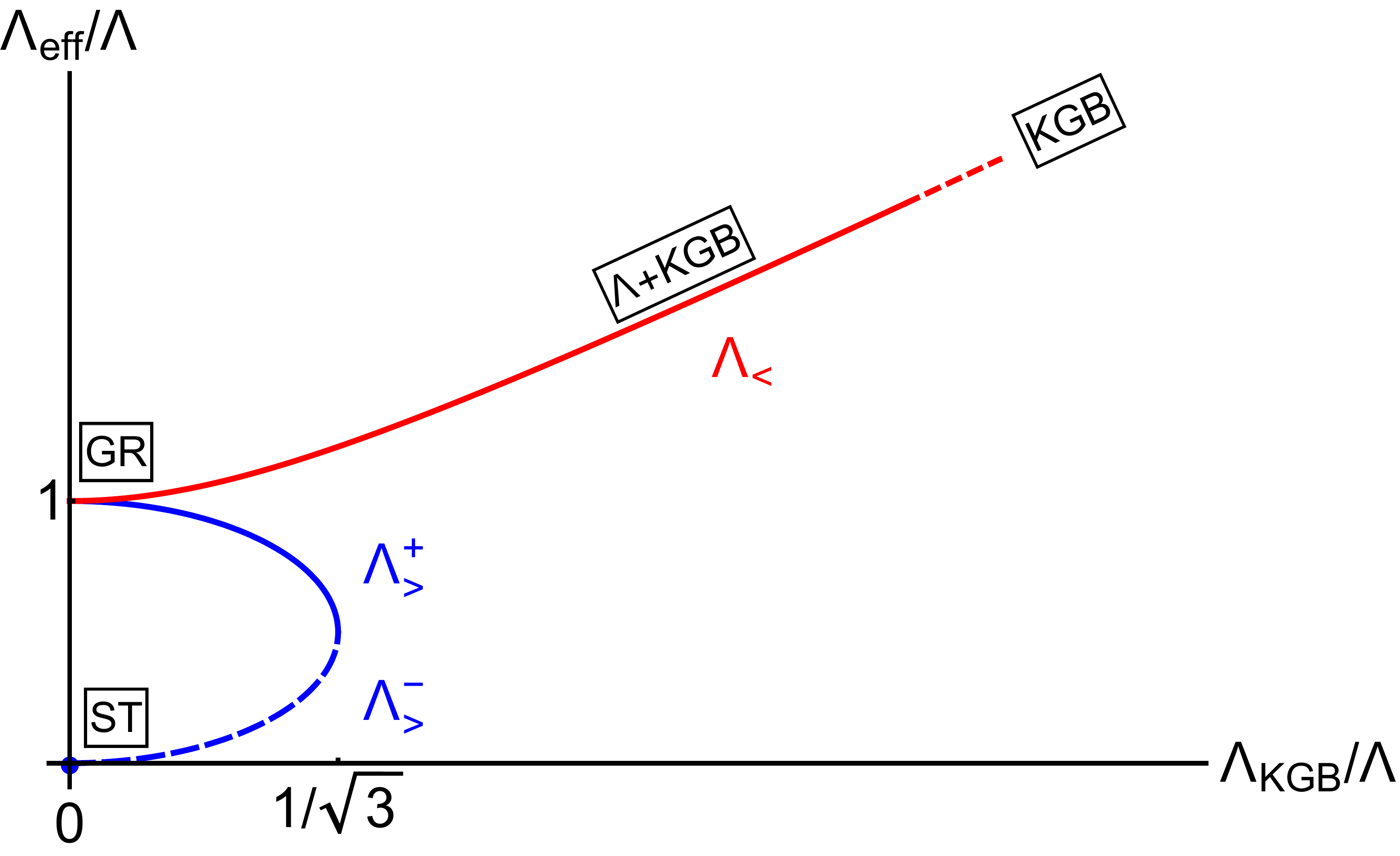}
\caption{Effective cosmological constant for black hole solutions in three dimensions. When $\Lambda_\mathrm{eff} \simeq \Lambda$, the solution behaves like GR. The other end of the red branch corresponds to a self-accelerated solution. The dashed blue branch represents a self-tuning (ST) solution, with $\Lambda_\mathrm{eff} < \Lambda$.
This graph remains nearly identical when we describe homogeneous cosmologies in four dimensions (see Sec. \ref{sec:homcosm}). This is why we put the KGB and $\Lambda+$KGB regimes immediately.}
\label{fig:lambda3d}
\end{center}
\end{figure}

The different branches are represented in Fig.~\ref{fig:lambda3d}. 
GR solutions are restored for $\Lambda_\mathrm{eff}= \Lambda$, in which case the presence of the cubic Galileon does not affect the metric. This limit can be attained either via $\Lambda_{>}^+$ or through $\Lambda_{<}$. The latter branch also attains, at the other end, the KGB limit as $\Lambda\rightarrow 0$ and we have self-accelerating solutions~\cite{Deffayet:2010qz} for $\Lambda_\mathrm{eff} \propto \Lambda_\mathrm{KGB}$.
The self-tuning solution corresponds to the lower branch $\Lambda_{>}^-$, where $\Lambda_\mathrm{eff} < \Lambda$. For this branch the scalar field partially screens the bare value $\Lambda$, yielding a cosmological constant of lesser magnitude.

\section{Black holes in 4D: Analytic approximations and asymptotics}
\label{Sec:4D}

Due to the more complex equations of motion in 4D (see below), we are not able to extract analytic black hole solutions. 
Therefore, we mostly resort to numerical integration of the equations of motion. 
It is possible, however,  to get some analytic insights about the solutions in different limits. 

For the ansatz~(\ref{eq:ansatz}), we get the following equations of motion, using the same notation as in the 3D case:
\begin{eqnarray}
\gamma \: q \: (r^4 h)' \dfrac{f}{h} \chi^2 - \gamma \: q^3 \: r^4 h' - 2\: \eta \: q \: r^4 h \chi &=& 0, \label{eq:tr}\\
\eta \: r^2 \left(\dfrac{f}{h} \chi^2 - q^2 \right) + 2 \: \zeta \: r f h' + 2\zeta \: h (f-1 + \Lambda \: r^2)  &=& 0, \label{eq:mix1}\\
\left(\dfrac{f}{h} \chi^2 - q^2 \right) \left[ \eta \: r^2 \sqrt{\dfrac{h}{f}} - \gamma \: \left(r^2 \sqrt{\dfrac{f}{h}} \chi \right)' \right] &=& 2 \: \zeta \: r h^2 \left(\sqrt{\dfrac{f}{h}} \right)',
\label{eq:mix2}
\end{eqnarray}
cf. Eqs.~(\ref{eq:tr3d})-(\ref{eq:mix23d}). 
In this section we will study solutions to the above system of three ODEs.

\subsection{Cosmological solutions}
\label{sec:homcosm}
We first consider homogeneous cosmological solutions for the model~(\ref{eq:action}). 
These homogeneous solutions describe the far away asymptotics for the black hole solutions we will search for with numerical integration later on. Let us note that the model (\ref{eq:action}) with $\Lambda =0$ has been studied previously in~\cite{EBGEF13} in the cosmological context. 
Here we extend the analysis of~\cite{EBGEF13} to include a non-zero $\Lambda$. 

A homogeneous solution written in spherical coordinates reads
\begin{equation}
\begin{split}
f(r) &= h(r) = \ 1 - \dfrac{\Lambda_\mathrm{eff}}{3} r^2,\\
\chi(r) &= \dfrac{\eta r}{3 \gamma},
\end{split}
\label{eq:cosmsol}
\end{equation}
where $\Lambda_\mathrm{eff}$ is to be determined from Eqs.~(\ref{eq:tr})-(\ref{eq:mix2}). Note also that, as in 3D, $q$ is not free; it is fixed to some particular value $q_0$, given in Eq. \ref{eq:q} below.
One can see the homogeneity of the previous solution by mapping Eqs.~(\ref{eq:cosmsol}) to a FLRW metric:
$$
\mathrm{d}s^2 = - \mathrm{d}\tau^2 + \text{e}^{2 H \tau} (\mathrm{d}\rho^2 + \rho^2 \mathrm{d}\Omega_D^2),
$$ 
thanks to the following coordinate transformation, \cite{EBGEF13}:
\begin{equation}
\begin{split}
\tau &= t + \sqrt{\dfrac{3}{4\: \Lambda_\mathrm{eff}}} \: \text{ln} \: \left( 1-\dfrac{\Lambda_\mathrm{eff}}{3} r^2 \right),\\
\rho &= r \: \text{e}^{-\sqrt{3/\Lambda_\mathrm{eff}} t} \: \left( 1-\dfrac{\Lambda_\mathrm{eff}}{3} r^2 \right)^{-1/2}.
\end{split}
\label{eq:FLRW}
\end{equation}
It is clear that the scalar field is homogenous when in the FLRW coordinates $\phi(\tau,\rho) = q_0 \: \tau$.
Equations~(\ref{eq:tr})-(\ref{eq:mix2}) then show that $q_0$ must be a solution of the equation
\begin{equation}
\dfrac{\eta^2}{3 \gamma^2 q_0^2} = \dfrac{2 \zeta \Lambda - \eta \: q_0^2}{2 \zeta}.
\end{equation}
There are two possible solutions for $q_0$ ($q$ appears only as $q^2$ in the equations of motion, so its overall sign is irrelevant):
\begin{equation}
q_0^\pm \equiv \left[\dfrac{\zeta \Lambda}{\eta} \pm \sqrt{\left(\dfrac{\zeta \Lambda}{\eta} \right)^2 - \dfrac{2 \eta \zeta}{3 \gamma^2}}\; \right]^{1/2},
\label{eq:q}
\end{equation}
corresponding to the two possible effective cosmological constants,
\begin{equation}
\Lambda_\mathrm{eff} = \dfrac{\eta^2}{3 \gamma^2 (q_0^\pm)^2} = \dfrac{2 \zeta\Lambda - \eta \: (q_0^\pm)^2}{2 \zeta}
\label{eq:Leff}
.\end{equation}
For $\Lambda = 0$, we find that 
\begin{equation}
\Lambda_\mathrm{eff} = \left(\dfrac{|\eta|^3}{6 \zeta \gamma^2}\right)^{1/2} = \Lambda_\mathrm{KGB}
\end{equation}
and $\eta$ has to be negative, in agreement with \cite{EBGEF13}.

To summarize, depending on the parameters of the model, the cosmological solutions are given by (\ref{eq:cosmsol}), (\ref{eq:q}), with the following values of the effective cosmological constant $\Lambda_\mathrm{eff}$, depending on the parameters of the model: 
\begin{equation}
\Lambda_\mathrm{eff}=\left\{
\begin{array}{r c l}
\Lambda_< &=& \dfrac{1}{2} \left(\Lambda + \sqrt{\Lambda^2+4\: \Lambda_\mathrm{KGB}^2}\right) \ \text{if} \ \eta <0,\\
\\
\Lambda_>^+ &=& \dfrac{1}{2} \left(\Lambda + \sqrt{\Lambda^2-4\: \Lambda_\mathrm{KGB}^2}\right) \ \text{if} \ \eta >0 \ \text{and} \ \Lambda > 2 \Lambda_\mathrm{KGB},\\
\\
\Lambda_>^- &=& \dfrac{1}{2} \left(\Lambda - \sqrt{\Lambda^2-4\: \Lambda_\mathrm{KGB}^2}\right) \ \text{if} \ \eta >0 \ \text{and} \ \Lambda > 2 \Lambda_\mathrm{KGB}.
\end{array}
\right.
\label{eq:Lambdaeff}
\end{equation}
Equation~(\ref{eq:Lambdaeff}) is similar to the 3D case, see Eq.~(\ref{eq:Lambdaeff3D}), apart from a numerical factor in front 
of $\Lambda_\mathrm{KGB}$ 
and a sign in $\Lambda_<$. 
Thanks to this similarity, Fig.~\ref{fig:lambda3d} also catches all the important features of the homogeneous solutions in the 4D case. 
Let us however point out the main differences between 3D and 4D cases. 
For a given set of parameters $\zeta$, $\eta$, $\gamma$ and $\Lambda$, the three dimensional solution presented in Sec.~\ref{Sec:3D}
includes a family of black holes, with a free mass parameter $M$. 
Besides, the spacetime is asymptotically anti de Sitter ($\Lambda_\mathrm{eff} <0$). 
On the other hand, in 4D, 
the solutions (\ref{eq:cosmsol}), (\ref{eq:q}) and (\ref{eq:Lambdaeff}) describe homogeneous de Sitter spacetimes.

\subsection{Test field limit}
\label{Ssec:test}

Before solving the full system of equations~(\ref{eq:tr})-(\ref{eq:mix2}) in the case of black holes, it is instructive to look into a particular limit, 
when the scalar field does not backreact onto the metric (the ``test field'' approximation). 
Formally this approximation can be obtained by setting $\eta = \epsilon \: \eta_0$ and $\gamma = \epsilon \: \gamma_0$, and then letting $\epsilon \rightarrow 0$ in 
the equations~(\ref{eq:tr})-(\ref{eq:mix2}). 
It is then easy to note that the metric is determined solely by the Einstein-Hilbert part of the action. 
The Schwarzschild-de Sitter metric with cosmological constant $\Lambda$ solves the two equations, Eqs.~(\ref{eq:mix1}), (\ref{eq:mix2}) in this limit. 
The third equation~(\ref{eq:tr}), the $(t r)$ metric equation, decouples from~Eqs.~(\ref{eq:mix1}), (\ref{eq:mix2}) 
and gives the scalar field equation on a fixed background metric. Explicitly, 
\begin{equation}
f(r) = h(r) = 1- \dfrac{\mu}{r} - \dfrac{\Lambda}{3}\: r^2,
\end{equation}
where $\mu$ is a mass parameter. Plugging the above expression into the $(t r)$ component of the metric equation yields,
\begin{equation}
\chi = \dfrac{\eta}{\gamma} \dfrac{r h \pm \sqrt{\Delta}}{(4\: h+ r h')},
\label{chitest}
\end{equation}
where 
\begin{equation}
\begin{split}
\Delta (r)  &=  \mu ^2+r^4 \left(\frac{4 \gamma ^2 \Lambda ^2 q^2}{3 \eta ^2}-\frac{2 \Lambda }{3}\right)+r^2 \left(1-\frac{8 \gamma ^2 \Lambda  q^2}{3 \eta ^2}\right)
\\
 &-  \frac{3 \gamma ^2 \mu ^2 q^2}{\eta ^2 r^2}+\frac{4 \gamma ^2 \mu  q^2}{\eta ^2 r}+ \frac{\Lambda ^2 r^6}{9} +\frac{2 \Lambda  \mu  r^3}{3}-2 \mu  r .
\end{split}
\end{equation}
Depending on the parameters of the Lagrangian, $\Delta (r)$ may become negative for some range of $r$, 
rendering the scalar field imaginary. 
One can check however, that in the case of physical interest, $1/\sqrt{\Lambda} \gg \mu$, 
$\Delta$ can be positive everywhere outside the horizon by requiring that
\begin{equation}
\left(\dfrac{\gamma q}{\eta} \right)^2 < \dfrac{1}{3 \Lambda}
.\end{equation}
The scalar field becomes imaginary for $r \lesssim 3/4 \mu$, i.e. in the interior region of the horizon. 

At this point we would like to make a remark on the physical meaning of our ansatz~(\ref{eq:ansatz}). 
Taking into account that the scalar field is time-dependent, the choice of a static spacetime metric is a non-trivial assumption. 
Indeed, although this ansatz ``passes through'' the equations of motion, leaving ordinary differential equations (instead of PDEs), 
the requirement of spacetime staticity implies no flux onto the black hole.
This situation is clearly uncommon in the case of ``usual'' matter~\cite{Babichev:2005py} and scalar fields~\cite{Babichev:2004yx,Babichev:2006vx}, 
when matter starts falling into a black hole, 
thus rendering a non-zero matter flow. The matter flow inevitably leads to a non-static metric.

A non-static ``accreting'' solution can also be found for the Lagrangian~(\ref{eq:action}). Indeed, in~\cite{Babichev:2010kj},
a process of accretion of Galileon onto a static spherically symmetric black hole was studied, where the backreaction of the scalar field on the black hole was neglected ---
exactly the situation we consider in this subsection, the test approximation. The key difference of the accreting solution in~\cite{Babichev:2010kj} from our solution~(\ref{chitest}) 
is an integration constant, which vanishes for the solution presented here.
As we discussed above, the solution in the test field approximation follows from the $(tr)$ Einstein equation~(\ref{eq:tr}), 
which is equivalent to the equation $J^r=0$. This last equation can be obtained from the scalar field equation~(\ref{Jcons}) by integrating along the radial coordinate and 
setting to zero the integration constant. 
In the case of accretion, this integration constant (the primary hair of the black hole) is not set to zero; instead, it is chosen in such a way that the solution for the scalar field describes a so called transonic flow,
so that it is smooth and free of singularities (at least for radii larger than the radius of the sound horizon). 

The above considerations thus suggest that, when backreaction is taken into account, other solutions --- with non-zero flux --- may exist. 
Therefore, the ansatz (\ref{eq:ansatz}) is not unique, but rather it corresponds to a special case of zero Galileon flow. 

\subsection{Asymptotic behavior at small and large \texorpdfstring{$r$}{}}
\label{sec:analytic}



Solving the system of equations (\ref{eq:tr})-(\ref{eq:mix2}) near the origin, $r\to 0$, we find the following asymptotic behavior,
\begin{equation}
\label{as1}
\begin{split}
h(r) &\simeq - b\: r^{-4} + c \: r^{-8/3}, 
\\
f(r) & \simeq - \dfrac{1}{3} + a \: r^{4/3}, 
\\
\chi(r) &\simeq d\:  r^{-13/3}, 
\end{split}
\end{equation}
%
%
where $a, b, c$ and $d$ depend of the parameters of the theory and are fixed by the field equations (their exact expressions are not interesting for us here). Note that unlike GR black holes, the $f(r)$ component of the metric is finite at the origin. Furthermore, one can actually show analytically that the behavior of the solutions near the black hole singularity depends only on the radial part of the scalar field and not on the time dependent part. Indeed, imposing a static $(q=0)$ scalar field, and further setting $\eta=0$, one can find an exact solution for all $r$ which has the same behavior as (\ref{as1}) in the $r\rightarrow 0$ region. Therefore, we can also conclude that, for $r\to 0$, the leading order behavior of the solution is determined by the higher-order Galileon term  $\Box \phi\: (\partial \phi)^2$, rather than by $(\partial \phi)^2$ or the $\Lambda$-term. This is expected as, close to the singularity, the higher order DGP term contains in total more derivatives than the $\eta$ and $\Lambda$ terms.
The numerical integration presented below confirms the behavior (\ref{as1}), see in particular  Figs.~\ref{fig:pureDGP} and \ref{fig:bhDGP}.

We now look for the large $r$ asymptotic behavior of the solution to Eqs. (\ref{eq:tr})-(\ref{eq:mix2}). We assume that, at spatial infinity, the solution has the following power expansion in $1/r$:
\begin{equation}
h(r) = \sum\limits_{n=-2}^\infty \dfrac{c^{(n)}_h}{r^n}, \;\;
f(r) = \sum\limits_{n=-2}^\infty \dfrac{c^{(n)}_f}{r^n}, \;\;
\chi(r) = \sum\limits_{n=-1}^\infty \dfrac{c^{(n)}_\chi}{r^n},
\end{equation}
with a de Sitter like behavior  at large $r$, i.e. $c^{(-2)}_h = c^{(-2)}_f$. Then, the asymptotic expansion reads
\begin{equation}\label{hfcfar}
\begin{split}
h(r) &= - \dfrac{\Lambda_\mathrm{eff}}{3} \: r^2 + 1 + \mathcal{O} \left(\dfrac{\mu}{r}\right),
\\
f(r) &= - \dfrac{\Lambda_\mathrm{eff}}{3} \: r^2 + c^{(0)}_f + \mathcal{O} \left(\dfrac{\mu}{r}\right),
\\
\chi(r) &= \dfrac{\eta r}{3 \gamma} + \dfrac{c^{(-1)}_\chi}{r} + \mathcal{O} \left(\dfrac{\eta \mu}{\Lambda_\mathrm{eff} \gamma r^2}\right),
\end{split}
\end{equation}
where $c^{(0)}_f$ and $c^{(-1)}_\chi$ are particular functions of the Lagrangian parameters (we do not give the exact expression for $c^{(0)}_f$ and $c^{(-1)}_\chi$ here, since they are cumbersome), 
and $\mu$ is a free integration constant. 
Note that $\mu$ is related to the black hole mass, and indeed we find that 
its order of magnitude should be the Schwarzschild radius of the black hole.
It is important to stress that the metric in the expansion~(\ref{hfcfar}) asymptotically approaches the metric of the homogeneous cosmological solution, 
since $\Lambda_\mathrm{eff}$ in~(\ref{hfcfar}) is given by~(\ref{eq:Leff}).   
Note that in this expansion, the velocity parameter $q$ remains arbitrary; it may not coincide with $q_0$, which is fixed by the cosmological solution. 

The question then arises whether the asymptotic solution~(\ref{hfcfar}) is homogeneous, since in the time-dependent part of the scalar field enters an arbitrary velocity $q$, which does not necessarily match the cosmological solution.
To check the homogeneity of the scalar field, we explicitly find the solution for $\phi$ by integration of~(\ref{hfcfar}):
\begin{equation}
\phi(t,r) \underset{r \rightarrow \infty}{=} q\: t - \dfrac{\eta}{\Lambda_\mathrm{eff} \gamma} \: \text{ln} \left(\sqrt{\dfrac{\Lambda_\mathrm{eff}}{3}} r \right) 
+ \mathcal{O}\left(\dfrac{q\: \Lambda_\mathrm{eff}}{r} \right),
\end{equation}
and then by the change to Friedmann coordinates we find:
\begin{equation}
\phi(\tau , \rho) \underset{\rho \rightarrow \infty}{=} q_0\: \tau + (q_0-q) \sqrt{\dfrac{3}{\Lambda_\mathrm{eff}}} \: \text{ln} \left(\sqrt{\dfrac{\Lambda_\mathrm{eff}}{3}} \rho \right) + \mathcal{O}\left(\dfrac{q\: \Lambda_\mathrm{eff}}{\rho \mathrm{e}^{\tau \sqrt{\Lambda_\mathrm{eff}/3}}} \right)
\label{eq:corrqq0}
.\end{equation}
This indeed proves that $\phi(\tau , \rho)$ is asymptotically homogeneous, even though it contains a slowly decaying inhomogeneous part at large $\rho$\footnote{It may seem 
that the solution is always inhomogeneous because the logarithmic part does not disappear in the limit $\rho\to \infty$. However, 
one should keep in mind that the value of $\phi$ itself is not a physical observable, because of the shift symmetry of the problem. 
Only derivatives of $\phi$ enter equations of motion. One can easily conclude from~(\ref{eq:corrqq0}) that 
$\frac{\partial\phi}{\partial\rho} \sim \rho^{-1}$, which becomes negligible with respect to $\frac{\partial\phi}{\partial\tau}=q_0$.}. $\phi$ behaves like the associated cosmological solution from Sec. \ref{sec:homcosm} i.e., the one with $q = q_0$. The situation is very similar to the one observed analytically for the ``John'' Galileon term, \cite{damos}. There, it was shown that the self-tuing black hole solutions obtained for a particular $q_0$ were insensitive to the change of the velocity parameter $q$, and that the effective cosmological constant, essentially $q_0$, remained unchanged. This is important for the self-tuning mechanism, as it does not rely on some fine tuned parameter $q_0$, it is rather generic.

In the case $q=q_0$, the previous expansion gets simplified as follows:
\begin{equation}
\begin{split}
h(r) &= 1 - \dfrac{\mu}{r} - \dfrac{\Lambda_\mathrm{eff}}{3} \: r^2   + \mathcal{O} \left(\dfrac{\mu^2 \Lambda_\mathrm{KGB}^2}{\Lambda_\mathrm{eff}^2 (\Lambda_\mathrm{eff}^2+\Lambda_\mathrm{KGB}^2)} \dfrac{1}{r^6}\right),
\\
f(r) &= 1- \dfrac{\mu}{r}  - \dfrac{\Lambda_\mathrm{eff}}{3} \: r^2  + \mathcal{O} \left(\dfrac{\mu^2 \Lambda_\mathrm{KGB}^2}{\Lambda_\mathrm{eff} (\Lambda_\mathrm{eff}^2+\Lambda_\mathrm{KGB}^2)} \dfrac{1}{r^4}\right),
\\
\chi(r) &= \dfrac{\eta r}{3 \gamma} + \dfrac{3 \gamma q_0^2 \mu}{2 \eta r^2} + \mathcal{O} \left( \dfrac{q_0 \mu^2 \Lambda_\mathrm{eff}^{1/2}}{(\Lambda_\mathrm{eff}^2+\Lambda_\mathrm{KGB}^2)} \dfrac{1}{r^5}\right),
\end{split}
\end{equation}
where $\Lambda_\mathrm{KGB}$, $q_0$ and $\Lambda_\mathrm{eff}$ are given correspondingly in (\ref{LambdaKGB}), (\ref{eq:q}) and (\ref{eq:Leff}), and 
$\mu$ is a free constant.
Here we see effectively the important role played by the time dependent part of the scalar field which determines the asymptotic behavior of the black hole solution as well as the modified value of the effective cosmological constant. 
The asymptotic solution for $\phi$ in Friedmann coordinates reads, in this case,
\begin{equation}
\phi(\tau , \rho) \underset{\rho \rightarrow \infty}{=} q_0 \: \tau + \mathcal{O} \left( \dfrac{q_0 \: \mu}{\text{e}^{3 \: \tau \sqrt{\Lambda_\mathrm{eff}/3}} \Lambda_\mathrm{eff}^{3/2} \rho^3} \right).
\label{eq:corrbh}
\end{equation}
Note the much faster decay of the inhomogeneous part, $\rho^{-3}$, in the case $q=q_0$ with respect to the case $q\neq q_0$ (\ref{eq:corrqq0}).

\section{Black holes in 4D: Numerical integration}
\label{sec:Numerics}
In this section, we will perform the numerical integration of the system of ODEs~(\ref{eq:tr})-(\ref{eq:mix2}), which is a consequence of the equations of motion of the model~(\ref{eq:action}) with the ansatz~(\ref{eq:ansatz}).
It is convenient to introduce dimensionless quantities in order to integrate numerically this system of equations. 
The theory~(\ref{eq:action}) contains four dimensionful parameters $\zeta, \eta, \gamma, \Lambda$. Besides, the ansatz for the scalar field 
has an extra dimensionful quantity $q$. Thus, in total there are five dimensionful parameters, which have to be combined in a number of dimensionless quantities.

Let us define first the dimensionless radius $x = r/r_0$ with some length scale $r_0$ (which we later choose to be the black hole radius). 
Then we define the three dimensionless constants as combinations of the parameters of the Lagrangian, the velocity $q$, and the length scale $r_0$ as follows:
\begin{equation}\label{eq:dlessconst}
\alpha_1 = - \dfrac{\gamma q}{r_0 \eta}, \;\; \alpha_2 = - \dfrac{\eta q^2 r_0^2}{\zeta},\;\; \alpha_3 = \Lambda r_0^2.
\end{equation}
%
%
Finally, we consider the dimensionless function $\chi/q$, i.e. the scalar field is measured in units of $q$. The equations of motion~(\ref{eq:tr})-(\ref{eq:mix2}) then can be rewritten in terms of the dimensionless quantities:
\begin{eqnarray}
\label{eq:tralpha}
\alpha_1\: (x^4 h)' \: \dfrac{f}{h} \left(\dfrac{\chi}{q}\right)^2 + 2\: x^4 h \left(\dfrac{\chi}{q}\right) - \alpha_1 \: x^4 h' &=& 0,\\
\label{eq:mix1alpha}
\alpha_2 \: x^2 \left[1 - \dfrac{f}{h} \left(\dfrac{\chi}{q}\right)^2 \right] + 2 \: x f h' + 2\: h (-1+f+ \alpha_3 \: x^2) &=& 0,\\
\left[1 - \dfrac{f}{h} \left(\dfrac{\chi}{q}\right)^2 \right] \left[\alpha_2 \: x^2 \: \sqrt{\dfrac{h}{f}} + \alpha_1 \: \alpha_2 \: \left(x^2 \sqrt{\dfrac{f}{h}} \dfrac{\chi}{q} \right)' \right] &=& 2 \: x h^2 \left(\sqrt{\dfrac{f}{h}} \right)',
\label{eq:mix2alpha}
\end{eqnarray}
where a prime throughout this section denotes a derivative with respect to the dimensionless radius $x$.
Henceforth, we choose the length scale $r_0$ to be the Schwarzschild radius of the black hole i.e., in terms of $x$ the black hole horizon is at $x=1$.

Note that the first two equations~(\ref{eq:tralpha}) and~(\ref{eq:mix1alpha}) are the algebraic equations on $f$ and $\chi$.  
Thus we can resolve Eqs.~(\ref{eq:tralpha}) and~(\ref{eq:mix1alpha}) to find $f$ and $\chi$ in terms of $h$ and $h'$. By substituting the obtained expressions in 
the third equation of the system, Eq.~(\ref{eq:mix2alpha}), we arrive to a second order ODE on $h$. 
To find the unique solution, two boundary conditions should be supplemented. 
We impose one boundary condition at the black hole horizon: we require that the radial function $h$ vanishes at $x = 1$ (which can be simply thought as a definition of the black hole horizon). 
As a second boundary condition, we specify (arbitrarily) the derivative of $h$ at the point $x =1$, $h'|_1$.
By integration from $x=1$ to large $x$, we then select such $h'|_1$ that the solution at large $x$ has a desired cosmological asymptotic behavior, 
in other words, we use the numerical {\it shooting} method. 

\subsection{The case \texorpdfstring{$\eta=\Lambda=0$}{}}
\label{sec:eta0Lambda0}

First, we consider the case of vanishing $\eta$ and $\Lambda$. The action (\ref{eq:action}) in this case contains only 
the Einstein-Hilbert and the cubic Galileon terms. 
The only relevant dimensionless parameter is $\alpha_1 \cdot \alpha_2$. 
In the absence of a black hole, the corresponding cosmological solution is Minkowski spacetime, represented by the blue dot at the origin in Fig.~\ref{fig:lambda3d}.

\begin{figure}[ht]
\begin{center}
\includegraphics[width=10cm]{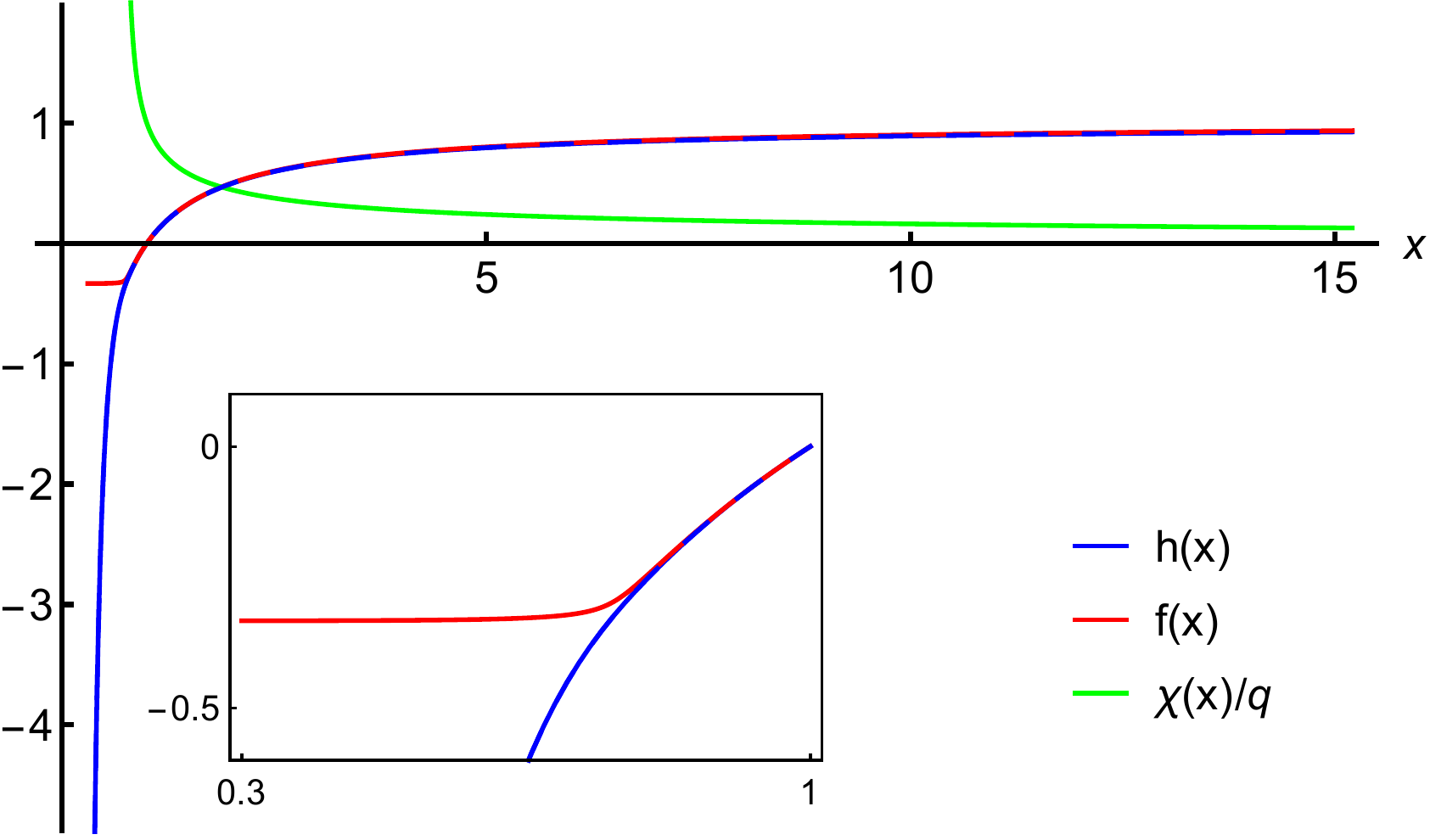}
\caption{Asymptotically flat black hole in the  $\eta=0$, $\Lambda=0$ case. For this solution,  $\alpha_1 \cdot \alpha_2 = 10^{-3}$, and the solution stops at $x=0.30$ for a numerical precision of 14 digits. The zoomed plot shows that there is no cusp in $f$.}
\label{fig:pureDGP}
\end{center}
\end{figure}

Solving numerically the system of equations (\ref{eq:tralpha})-(\ref{eq:mix2alpha}), we get asymptotically flat black holes, as shown in Fig. \ref{fig:pureDGP}.
For general boundary condition of the ODEs, $f$ and $h$ approach different constants at infinity. 
However, they can be matched by adjusting the numerical value of the derivative of $h$ at the location of the event horizon, $h'|_1$. 
The numerical solutions are always well-behaved in the direction of increasing $r$ (note that we perform numerical integration from the event horizon).
We find, however, that when the numerical precision is increased, 
the numerical integration cannot be continued below some radius {\it inside} the horizon, because the numerical code breaks down there. 
It should be stressed that this is a generic feature of all the simulations we carried out (see below), and not specific to the $\eta=\Lambda=0$ case. 
We could not conclude on the origin of this numerical singularity: it can be either a numerical artefact or a physical pathology at that point. 
However, the presence of a physical singularity inside the horizon in the test field limit, as we have shown in Sec.~\ref{Ssec:test}, suggests that 
the break down of the numerical code indeed signals about a singular behavior of a solution, rather than a numerical 
glitch.

\subsection{Generic case}

\begin{figure}[ht]
\centering
   \begin{subfigure}[b]{0.75\textwidth}
   \includegraphics[width=1\linewidth]{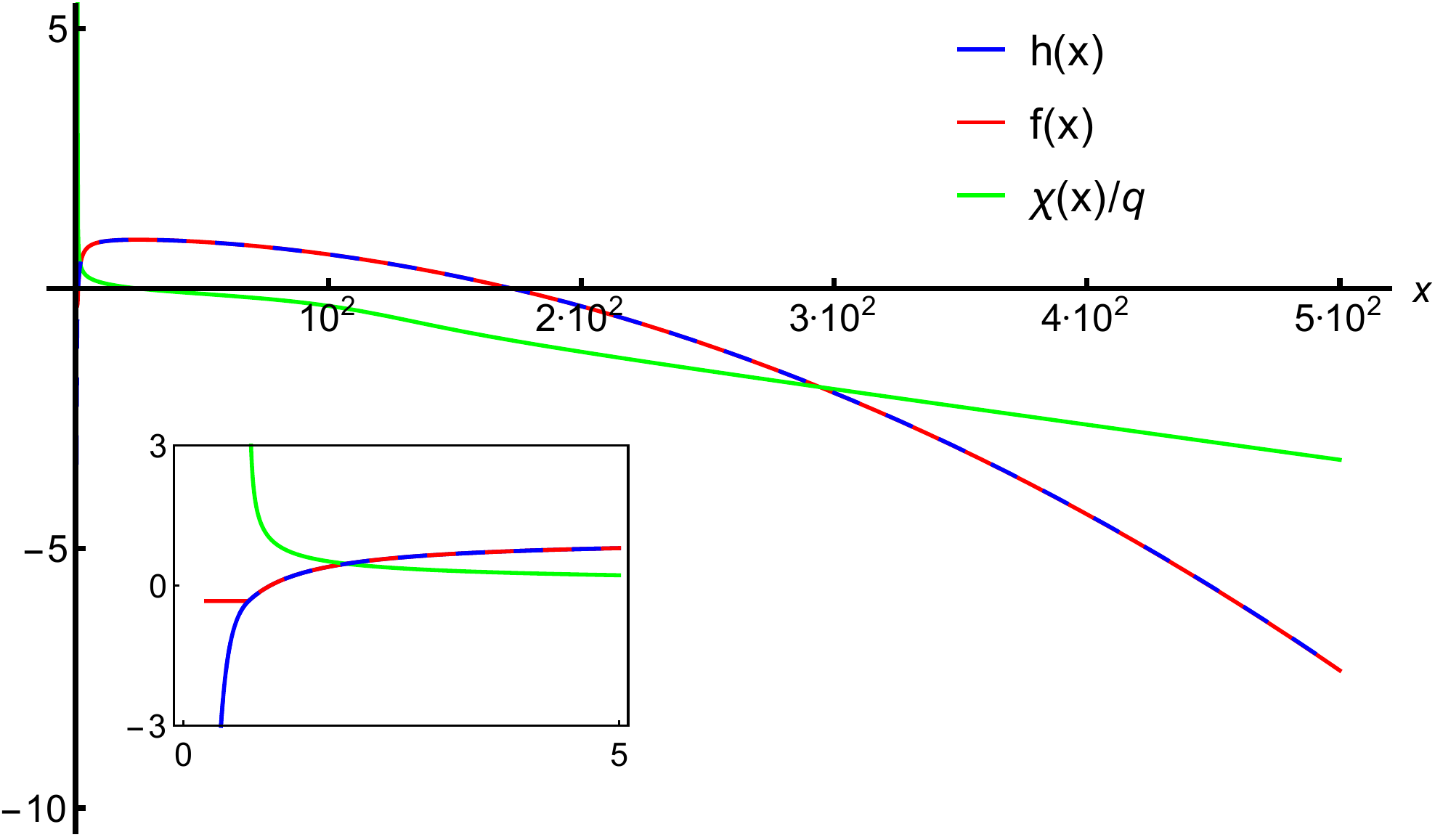}
   \caption{$\eta<0$}
   \label{fig:bhDGPetaneg} 
\end{subfigure}
\begin{subfigure}[b]{0.75\textwidth}
   \includegraphics[width=1\linewidth]{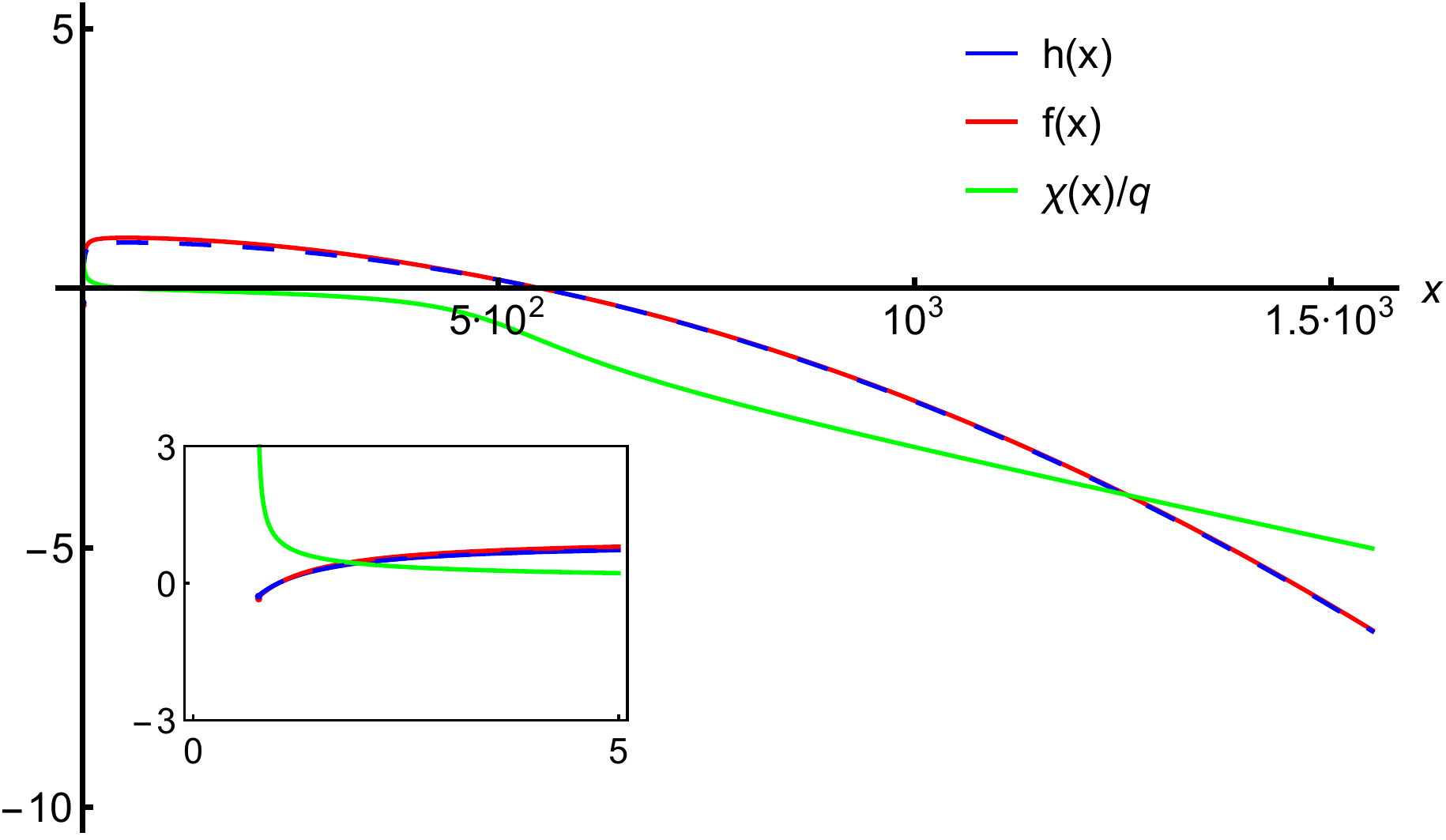}
   \caption{$\eta>0$}
   \label{fig:bhDGPetapos}
\end{subfigure}
\caption{(a) Typical black hole in a de Sitter Universe for the cubic Galileon. The parameters of this solution are $\alpha_1 = 50$, $\alpha_2 = 2.5 \cdot 10^{-7}$ and $\alpha_3 = 10^{-4}$. For this choice of parameters, $\eta<0$, the velocity is $q \simeq 0.87 \: q_0$ and the bare cosmological constant $\Lambda$ is about 25 times greater than the KGB one, $\Lambda_\mathrm{KGB}$. This solution is therefore in the $\Lambda_<$ branch, close to the GR regime (see Fig. \ref{fig:lambdanum}). The framed plot shows a zoom on the black hole region. (b) Another solution sitting in the $\Lambda_>^+$ branch, with $\alpha_1 = 10^2$, $\alpha_2 = -3 \cdot 10^{-7}$ and $\alpha_3 = 10^{-5}$; $q \simeq .53 \: q_0$ and $\Lambda \simeq 5 \Lambda_\mathrm{KGB}$.}
\label{fig:bhDGP}
\end{figure}

In this subsection, we consider general nonzero values of  $\eta$ and $\Lambda$. 
Typical behavior of such solutions are presented in Fig.~\ref{fig:bhDGP}.
In contrast to the case $\eta=\Lambda=0$, the asymptotic solutions are no longer flat, we are seeking de-Sitter asymptotic, 
according to our study above.

Let us comment at this point on some details of the numerical solutions we present here. 
We use the shooting method, starting from the location of the black hole horizon, i.e. $h|_1=0$ in the rescaled quantities. 
The value of the derivative $h'|_1$ is not, however, fixed by the condition at the event horizon. 
Whenever we find some numerical solution, $f$ and $h$ always behave like $r^2$ at large $r$, but with $f/h \neq 1$ asymptotically, in general. 
We therefore use the freedom of choosing  $h'|_1$ so that $f=h\sim r^2$ at $r\to \infty$.
This is, however, only possible to do for some range of $q$ (assuming the parameters of the Lagrangian are fixed) 
such that $q$ does not deviate too much from the value $q_0$.
In this case, there is a unique choice of $h'|_1$ so that $f$ and $h$ coincide at large $r$. 
On the contrary, for the values of $q$ that are far from $q_0$ it is impossible to do so, whichever boundary condition we choose.

In what follows, we will focus on the solutions for which $f/h=1$ asymptotically at large $r$, and we will discard other solutions. 
These numerical solutions have de-Sitter-like asymptotic behavior,
\begin{equation}
\label{numass}
\begin{split}
h(r) &\underset{r \rightarrow \infty}{\sim} f(r) \underset{r \rightarrow \infty}{\sim} - C_1 r^2,\\
\chi(r) &\underset{r \rightarrow \infty}{\sim} - C_2 r,
\end{split}
\end{equation}
with some positive constants $C_1$, $C_2$, c.f. Eq.~(\ref{hfcfar}).
In addition, we checked that the norm of the derivative of the scalar, $(\partial \phi)^2$, approaches a constant value at infinity, a further consistency check with the analytic cosmological solution (\ref{eq:cosmsol}). 


Depending on the choice of parameters and therefore of the particular branch, one may expect that all the black hole solutions fall in 
one of the three families, with a corresponding asymptotic value of  $\Lambda_\mathrm{eff}$, 
see the discussion in Sec.~\ref{sec:homcosm}. 
We were indeed able to find black hole solutions for both positive and negative $\eta$. 
For positive $\eta$, however,
which gives two cosmological branches, $\Lambda_>^\pm$, 
we only found numerical solutions which approach one of the branches, the $\Lambda_>^+$ one.

\begin{figure}[ht]
\begin{center}
\includegraphics[width=12cm]{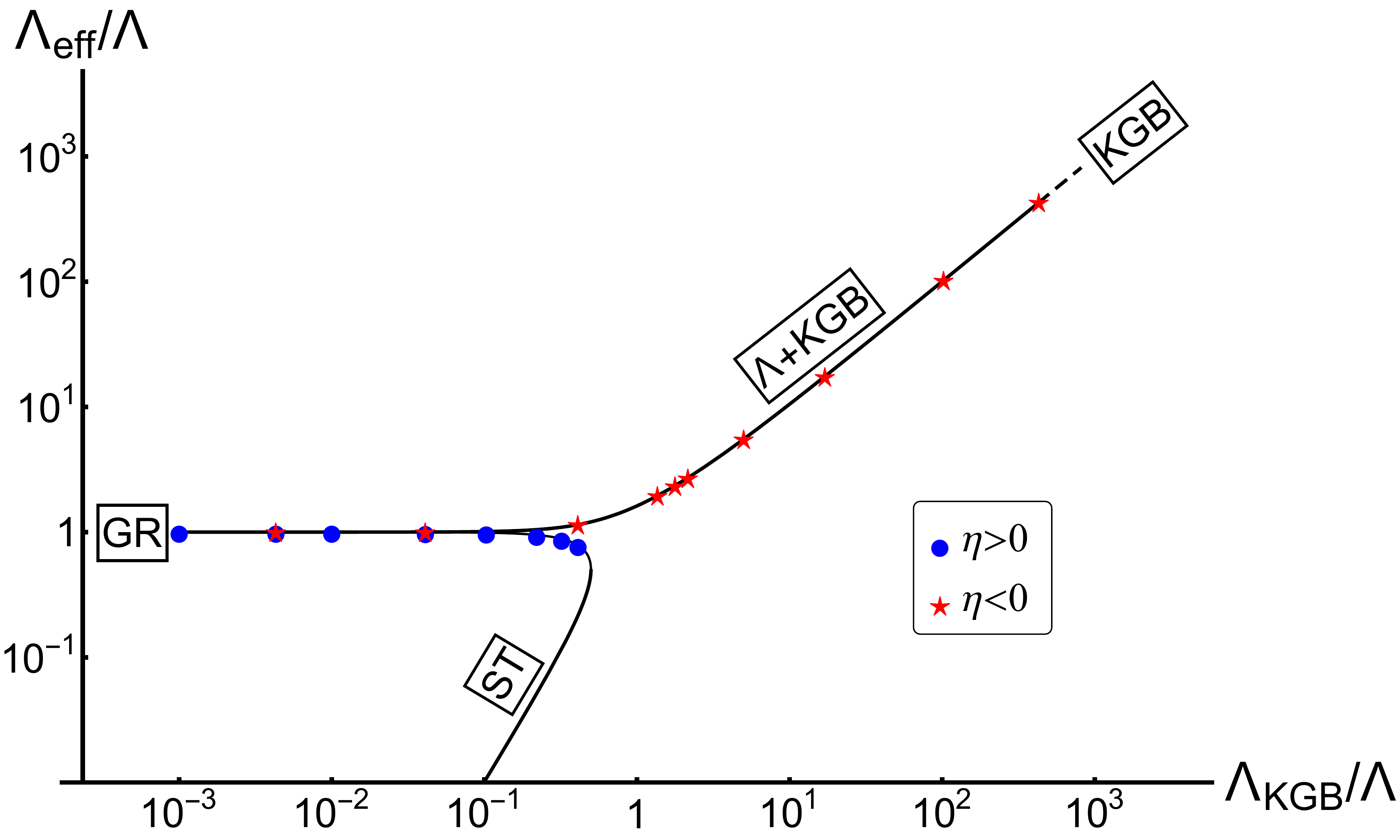}
\caption{Comparison between the far away metric of black hole solutions and their associated cosmological solutions. The black solid line is the value of $\Lambda_\mathrm{eff}/\Lambda$ expected from cosmology as a function of $\Lambda_\mathrm{KGB}$. The blue dots (red stars) represent the values of the same quantity obtained from numerical simulations with $\eta>0$ ($\eta<0$). We see a perfect agreement between the numerical results and the theoretical prediction.}
\label{fig:lambdanum}
\end{center}
\end{figure}

For a set of parameters $\zeta$, $\eta$, $\Lambda$, $\gamma$, the cosmological solution is given by (\ref{eq:cosmsol}) with $\Lambda_\mathrm{eff}$ 
along one of the branches of Eq.~(\ref{eq:Lambdaeff}) and $q_0$ given by (\ref{eq:q}).
Remarkably, we find that all numerical solutions for a fixed set $\zeta$, $\eta$, $\Lambda$, $\gamma$ 
asymptotically approach the $\Lambda_<$ cosmological solution for $\eta<0$ and 
the $\Lambda_>^+$ cosmological solution for $\eta>0$.
This means that the constants $C_1$, $C_2$ in Eq.~(\ref{numass}) are respectively $\Lambda_\mathrm{eff}/3$ and $\eta/(3 \gamma)$. In Fig.~\ref{fig:lambdanum}, we show the (normalized) cosmological constant which we read off our numerical solutions, versus the analytical results for the homogeneous cosmological solutions.

\begin{figure}[ht]
\begin{center}
\includegraphics[width=12cm]{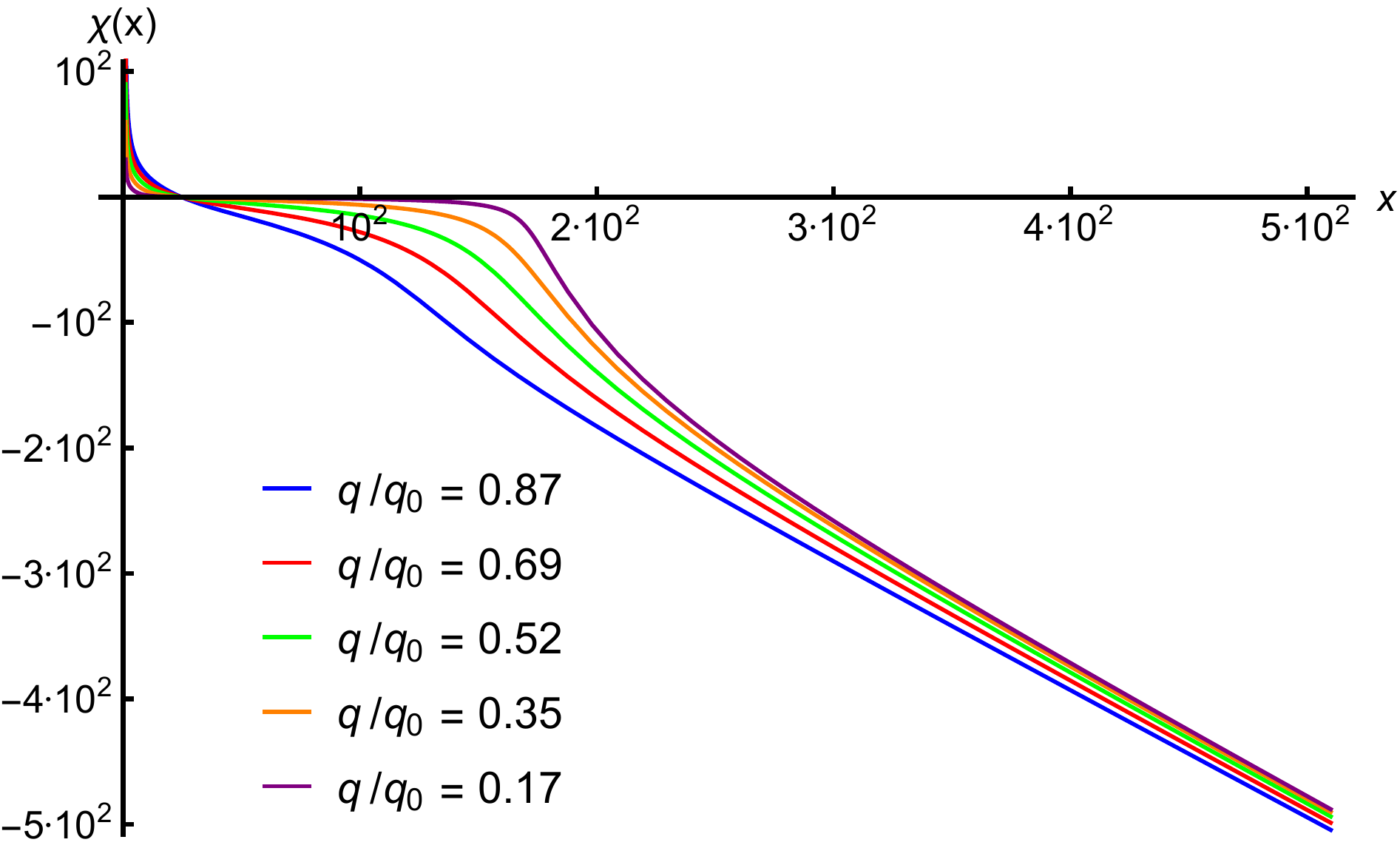}
\caption{The scalar field function $\chi$ for different values of the velocity $q$. The parameters of the Lagrangian are kept constant (they are the same as in Fig. \ref{fig:bhDGPetaneg}). Here, $\chi$ is measured in units of $\eta/(3 \gamma)$, which is the slope expected from Sec. \ref{sec:homcosm} far away. The solutions all behave identically far away from the black hole.}
\label{fig:primhair}
\end{center}
\end{figure}

It is important to stress here that the numerical values $C_1$, $C_2$ \textit{do not depend} on a particular value of $q$, 
which is a free parameter entering the scalar field ansatz and eventually the definition of dimensionless parameters 
$\alpha_i$ via (\ref{eq:dlessconst}). The value of $q$ determines the details of the black hole solutions, but not the far away behavior, as we expected from 
the discussion in~Sec.~\ref{sec:analytic}. Therefore, as shown in Fig. \ref{fig:primhair}, there exists a whole family of solutions parametrized by $q$ for a given set of parameters in the Lagrangian and, more importantly, for a given black hole mass. Indeed, in Fig. \ref{fig:primhair} for instance, the horizon location is kept fixed. The velocity parameter $q$ thus has the characteristics of primary hair.

It is worth mentioning that from the numerical solutions we have described above, with $f\sim h$ at large $r$, 
one can construct physically equivalent solutions with $f \neq h$ at infinity.
Indeed, changing the time parametrization as $t'= t/\sqrt{C}$, gives
\begin{equation}
\begin{split}
\phi(t',r) &= q \sqrt{C} t' + \displaystyle\int{\mathrm{d}r \: \dfrac{\chi(r)}{h(r)}},\\
\mathrm{d}s^2 &= - C h(r) \mathrm{d}t'^2 + \dfrac{\mathrm{d}r^2}{f(r)} + r^2 \mathrm{d}\Omega^2.
\end{split}
\end{equation}
instead of (\ref{eq:ansatz}).
Defining $\tilde{h} = C h$ and $\tilde{\chi} = C \chi$, we get back the old ansatz (\ref{eq:ansatz}) with $q \to q \sqrt{C}$.
Note that if the solution in the coordinates $(t,r)$ has $f\sim h$ at large $r$, the same solution in the coordinates $(t',r)$ has the asymptotic behavior 
$\tilde{h} \sim C f$.
In terms of dimensionless parameters, this corresponds to replacing $(\alpha_1, \: \alpha_2)$ by $(\sqrt{C} \alpha_1, \: C \alpha_2)$. 
It is clear, however, that all these solutions with arbitrary $C$ are physically equivalent.



\section{Conclusions}
\label{Sec:con}

In this paper we studied  black hole solutions in a subclass of Horndeski theory, 
which contains the Einstein-Hilbert term, a cosmological constant, the quadratic and the cubic Galileon terms~(\ref{eq:action}). 
The solutions we find can be interpreted as black holes immersed in a self-accelerated, flat or self-tuning universe, depending on the asymptotic behavior.
Due to the non-trivial scalar field profile, which results from the time-dependence of the scalar field, the black hole solutions 
do not coincide with those of GR. 

For the model (\ref{eq:action}), we assumed a time-dependent ansatz for the scalar field and a static ansatz for the metric~(\ref{eq:ansatz}) --- 
an approach which has been put forward in~\cite{Babichev:2013cya} to look for black hole solutions in the Galileon theory with the ``John'' term. 
We first studied the 3D case, where equations are slightly simpler and certain solutions  can be found analytically, see~(\ref{sol3D}).
The solutions~(\ref{sol3D}) feature the BTZ solution, with an important difference, however, that our solutions
contain a nontrivial scalar field, which effectively modifies the cosmological constant. 
The scalar field depends on the black hole mass, hence it corresponds to secondary hair. 

Unlike the 3D case, in 4D we were not able to integrate the full system of equations exactly in the case of spherical symmetry. 
It is nevertheless possible to study analytically various asymptotic regimes and some specific limits (Sec.~\ref{Sec:4D}),
obtaining insight about the full solutions. 

We performed numerical integration of the full system of equations for different parameter ranges, Sec.~\ref{sec:Numerics}.
We were able to find numerical solutions for a range of parameters of the theory, and of the scalar field velocity $q$ entering the scalar field ansatz~(\ref{eq:ansatz}). 
It is important to stress that solutions exist for a range of $q$ (for fixed parameters of the theory $\zeta$, $\eta$, $\Lambda$ and $\gamma$), 
which is a free parameter, independent on the mass of a black hole. 
At the same time, the asymptotic behavior at large distances is controlled by the fixed --- in terms of the Lagrangian parameters --- value $q_0$~(\ref{eq:q}), 
see a discussion in Sec.~\ref{sec:analytic}. 
Thus $q$ can be treated as a parameter corresponding to primary hair, 
since it determines the behavior of a solution at small distances, 
but at large radii the solution restores to the cosmological homogeneous configuration, independently on $q$. 
This interpretation of $q$ should be taken with the following reservation: 
the solution for $\phi$ in fact depends on $q$~(\ref{eq:corrqq0}) even asymptotically; however, 
the value of $\partial_\mu\phi$ approaches to the cosmological homogeneous solution (in Friedmann coordinates).
Also, we would like to note that the value of $q$ cannot be taken too different from its cosmologically defined value $q_0$: for 
$q$ deviating too much from $q_0$, we could not find any solutions.

There are several issues we left for future study. 
First of all, stability of the presented black hole solutions should be investigated in detail, namely, the solutions should be checked for 
possible ghost, gradient or tachyon instability. 
Also, the physical relevance of the ansatz~(\ref{eq:ansatz}) should be investigated further. Indeed, this ansatz describes a static time-independent metric, 
while the scalar field is time-dependent. By using this ansatz, we obtain a mathematically self-consistent system of ODE. 
However, it is not clear if such a configuration actually takes place during matter collapse. 
A corresponding numerical simulation of a time-dependent collapse might prove difficult because of at least two reasons:
scalar wave emission due to the absence of the Birkhoff theorem and caustic formation in Galileon theory~\cite{Babichev:2016hys}. This is however a very important open problem which we hope to study in detail in the near future. Furthermore, our analysis here for a cubic Galileon term, the DGP term, shows different behavior to the stealth GR like behavior observed in Galileon models including the ``John'' term~\cite{Babichev:2013cya}. 
It is therefore important to extend our analysis to include other terms in the action and to consider more general subclasses of Horndeski theory and beyond. 

 \acknowledgments
We are grateful to Gilles Esposito-Far\`ese and Mokhtar Hassaine for helpful discussions. The authors acknowledge financial support from the research program, Programme
national de cosmologie et galaxies of the CNRS/INSU, France. EB was supported in part by Russian Foundation for Basic Research Grant No. RFBR 15-02-05038. TM gratefully acknowledges support of the University Paris-Sud and Science and Technology Center in Ukraine.

\end{document}